\documentclass[revtex4]{emulateapj}
\usepackage{amsmath}
\usepackage{graphicx}
\usepackage{lscape}
\usepackage{epsfig}
\usepackage{natbib}
\bibpunct{(}{)}{;}{a}{}{,} % to follow the A&A style
\def\apjl{ApJ}%
\def\apjs{ApJS}%
\def\aap{A\&A}%
\shorttitle{CHARACTERIZING KIC\,10920273 AND KIC\,11395018}
\shortauthors{Dogan et al.}

\begin{document}

\title{CHARACTERIZING TWO SOLAR-TYPE $\emph{KEPLER}$ SUBGIANTS WITH ASTEROSEISMOLOGY: \\KIC\,10920273
AND KIC\,11395018}

\author{G. DO\u{G}AN$^{1, 2, 3}$, T. S. METCALFE$^{1, 3, 4}$, S. DEHEUVELS$^{3, 5}$, M. P. DI MAURO$^{6}$, P. EGGENBERGER$^{7}$,
O. L. CREEVEY$^{8,9,10}$, M. J. P. F. G. MONTEIRO$^{11}$, M.
PINSONNEAULT$^{3, 12}$, A. FRASCA$^{13}$, C. KAROFF$^{2}$, S.
MATHUR$^{1}$, S. G. SOUSA$^{11}$, I. M. BRAND{\~A}O$^{11}$, T. L.
CAMPANTE$^{11,18}$, R. HANDBERG$^{2}$, A.O. THYGESEN$^{2,14}$, K.
BIAZZO$^{15}$, H. BRUNTT$^{2}$, E. NIEMCZURA$^{16}$, T. R.
BEDDING$^{17}$, W. J. CHAPLIN$^{3, 18}$, J.
CHRISTENSEN-DALSGAARD$^{2, 3}$, R. A. GARC\'{I}A$^{3, 19}$, J.
MOLENDA-\.ZAKOWICZ$^{16}$, D. STELLO$^{17}$, J. L. VAN SADERS$^{3,
12}$, H. KJELDSEN$^{2}$, M. STILL$^{20}$, S. E. THOMPSON$^{21}$, and J.
VAN CLEVE$^{21}$} \affil{$^{1}$ High Altitude Observatory,
National Center for Atmospheric Research, P.O. Box 3000, Boulder, CO
80307, USA} \email{gulnur@ucar.edu} \affil{$^{2}$ Stellar Astrophysics Centre, Department of Physics and Astronomy,
Aarhus University, Ny Munkegade 120, DK-8000 Aarhus C, Denmark}
\affil{$^{3}$Kavli Institute for Theoretical Physics, Kohn Hall,
University of California, Santa Barbara, CA 93106, USA} \affil{$^{4}$ Space Science Institute,
Boulder, CO 80301, USA} \affil{$^{5}$Department of Astronomy, Yale
University, PO Box 208101, New Haven, CT 06520-8101, USA}
\affil{$^{6}$ INAF-IAPS, Istituto di Astrofisica e Planetologia
Spaziali, Via del Fosso del Cavaliere 100, 00133 Roma, Italy}
\affil{$^{7}$ Geneva Observatory, University of Geneva, Maillettes
51, 1290 Sauverny, Switzerland}

\affil{$^{8}$ Universit\'e de Nice, Laboratoire Cassiop\'ee, CNRS
UMR 6202, Observatoire de la C\^ote d'Azur, BP 4229, 06304 Nice
cedex 4, France} \affil{$^{9}$ IAC Instituto de Astrof\'{i}sica de
Canarias, C/ V\'ia L\'actea s/n, E-38200 Tenerife, Spain}

\affil{$^{10}$ Universidad de La Laguna, Avda. Astrof\'isico
Francisco S\'anchez s/n, 38206 La Laguna, Tenerife, Spain}

\affil{$^{11}$ Centro de Astrof\'{i}sica and DFA-Faculdade de
Ci\^{e}ncias, Universidade do Porto, Portugal} 

\affil{$^{12}$Ohio
State University, Department of Astronomy, 140 W. 18th Ave.,
Columbus, OH 43210, USA} 

\affil{$^{13}$ INAF, Osservatorio
Astrofisico di Catania, via S. Sofia, 78, 95123, Catania, Italy }

\affil{$^{14}$ Zentrum f\"ur Astronomie der Universit\"at Heidelberg,
Landessternwarte, K\"onigstuhl 12, 69117 Heidelberg, Germany}

\affil{$^{15}$ INAF - Osservatorio Astronomico di Capodimonte,
Salita Moiariello 16, 80131, Napoli, Italy} 

\affil{$^{16}$  Instytut
Astronomiczny, Uniwersytet Wroc\l{}awski, ul. Kopernika 11, 51-622
Wroc\l{}aw, Poland} 

\affil{$^{17}$  Sydney Institute for Astronomy
(SIfA), School of Physics, University of Sydney, NSW 2006,
Australia} 

\affil{$^{18}$ School of Physics and Astronomy,
University of Birmingham, Edgbaston, Birmingham, B15 2TT, UK}

\affil{$^{19}$ Laboratoire AIM, CEA/DSM -- CNRS - U. Paris Diderot --
IRFU/SAp, Centre de Saclay, 91191 Gif-sur-Yvette Cedex, France}

\affil{$^{20}$ Bay Area Environmental Research Institute / NASA Ames
Research Center, Moffett Field, CA 94035, USA} 

\affil{$^{21}$ SETI
Institute / NASA Ames Research Center, Moffett Field, CA 94035, USA}

\begin{abstract}

Determining fundamental properties of stars through stellar modeling
has improved substantially due to recent advances in
asteroseismology. Thanks to the unprecedented data quality obtained
by space missions, particularly CoRoT and $\textit{Kepler}$, invaluable information is extracted from the high-precision stellar oscillation frequencies, which provide very strong constraints on possible stellar
models for a given set of classical observations. In this work, we have characterized two relatively faint stars,
KIC\,10920273 and KIC\,11395018, using oscillation data from
$\textit{Kepler}$ photometry and atmospheric constraints from
ground-based spectroscopy. Both stars have very similar atmospheric
properties; however, using the individual frequencies extracted from
the $\textit{Kepler}$ data, we have determined quite distinct global
properties, with increased precision compared to that of earlier
results. We found that both stars have left the main sequence and
characterized them as follows: KIC\,10920273 is a one-solar-mass star
($M=1.00\pm0.04M_{\odot}$), but much older than our Sun
($\tau=7.12\pm0.47$\,Gyr), while KIC\,11395018 is significantly more
massive than the Sun ($M=1.27\pm0.04M_{\odot}$) with an age close to
that of the Sun ($\tau=4.57\pm0.23$ Gyr). We confirm that the high lithium abundance reported for these stars
should not be considered to represent young ages, as we precisely
determined them to be evolved subgiants. We discuss the use of
surface lithium abundance, rotation and activity relations as potential age diagnostics.
\end{abstract}

\keywords{asteroseismology --- stars: evolution --- stars: fundamental parameters --- stars: individual (KIC\,10920273, KIC\,11395018) --- stars: solar-type}

\section{Introduction}

Classical modeling of single stars mostly relies on fitting to the
atmospheric properties obtained through spectroscopic and/or photometric
observations, such as effective temperature, surface gravity, and
elemental abundances. This yields a large number of possible models
covering a wide range of values for the fundamental properties,
particularly in the absence of independent radius and luminosity
measurements. Asteroseismology has been revolutionizing stellar
modeling as a result of the precise and accurate inferences on
stellar structure that have been made possible by a new generation
of asteroseismic observations. Stellar fundamental properties, particularly mass and radius, can be determined to a few percent uncertainty,
even when only the average seismic parameters are used as
additional constraints, with the precision on age determination being as good as $\sim$20\%. The precision and accuracy of these
properties increase further when the individual oscillation
frequencies are used (see, e.g., \citealt{mathur2012} for a comparison between using average and individual seismic quantities as constraints in the modeling of a large sample of $\textit{Kepler}$ stars). Moreover, the individual frequencies allow us to
obtain information about the stellar interiors. 

Asteroseismology has proved very effective in constraining the
stellar age and the evolutionary stage with the help of
specific features seen in the oscillation spectra. Mixed modes are
particularly important in this regard. As a star evolves, the frequencies of the p modes decrease due to increasing stellar size, while
the g-mode frequencies increase.  By the time the star moves off the main
sequence, the g- and p-mode trapping cavities are closer to
each other, which results in the interaction of the two types of modes as they go through ``avoided crossings". The modes affected by this interaction are referred to as mixed modes due to having
g-mode characteristics in the deep interior, and p-mode characteristics near the surface, of the star (see \citealt{osaki75} and \citealt{aizenman77} for an
introductory discussion). They are sensitive to the central conditions and hence encode information from the core, where the chemical composition changes due to the nuclear reactions driving the evolution of the star. Since the
timescales of avoided crossings are very small compared to the stellar
evolutionary timescale, mixed modes provide very strong constraints on
the stellar age (see, e.g., \citealt{Deheuvels2011, metcalfe10ApJ, Benomar2012} for recent analyses).

$\textit{Kepler}$ is a space telescope with a diameter of  0.95\,m that has
been providing high-quality photometric data since the beginning of
its operations in May 2009 (see, e.g.,
\citealt{borucki2010,koch10,chaplin2010, chaplin2011}). The
mission's primary objective is to search for Earth-sized planets
through the transit method. Asteroseismology is being used to
characterize a subsample of stars, some of which host planets. $\textit{Kepler}$ monitors
more than 150,000 stars, and $\sim$2000 of these were selected to be monitored for one month each in
short-cadence mode (58.9\,s integrations) during the first 
$\sim$10 months of the mission \citep{Gilliland2010,chaplin2011}. Solar-like oscillations were detected in at least 500 of those survey stars \citep{chaplin2011}. A subsample ($\sim$190) of these have been monitored for more than 3 months, and precise determination of the oscillation properties has been completed for part of the sample \citep{Thiery2012}. Asteroseismology has been proving successful in determining their global properties and inferring their interiors (see, e.g, \citealt{metcalfe10ApJ,metcalfe2012, Creevey2012,  Deheuvels2012, mathur2012}).

KIC\,10920273 (kepmag=11.93\,mag, i.e., apparent magnitude as observed through the $\textit{Kepler}$ bandpass) and KIC\,11395018 (kepmag=10.76\,mag)
are among a handful of asteroseismic targets that were observed continuously from the start of science operations. Consequently, extended timeseries were available from early in the mission, making
both stars attractive targets for asteroseismic analysis \citep{Campante11, Mathur11}. We also
acquired ground-based spectra in order to characterize these
stars. They are G-type stars with very similar spectroscopic
properties, especially $T_{\rm eff}$ and $\log g$ (see, Sect.\,2.1),
so it is difficult to discriminate between models for the two
stars using classical approaches. These approaches include matching
the position of the star in the Hertzsprung-Russell (H-R) diagram in
the form that shows luminosity versus effective temperature as the
star evolves, or alternatively in the $\log g$-$T_{\rm eff}$ diagram, given that the luminosity cannot be calculated
using the available observations.

We present the observational data employed to characterize our stars
in Section~2, our modeling approach in Section~3, and the results in
Section~4, while Section~5 provides a summary and conclusions.

\section{Observational constraints}
\subsection{Atmospheric properties}

\begin{figure}[h]
\begin{center}
\hspace{-5mm}\includegraphics[scale=0.35,angle=90]{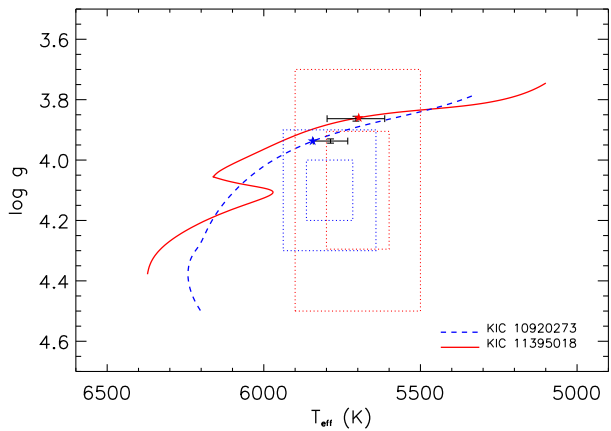}\\
\end{center}
\caption{ Log\,$g$-T$_{\rm eff}$ diagram for KIC\,10920273 and
KIC\,11395018. Surface gravity, $g$, is in cgs units. Spectroscopic
constraints given in Table~\ref{tbl_atm} are shown by 1- and
2-$\sigma$ error boxes (dotted; blue for KIC\,10920273 and red for
KIC\,11395018). Evolutionary tracks of two models indicated by star
symbols (SA1 and
BA1 from Tables\,\ref{scully} and \ref{boogie}) are plotted using the same color code. The points with error bars represent the weighted means and the
standard deviations of the asteroseismic determinations (see Tables\,\ref{scully} and \ref{boogie}).}
\label{HR_AMP}
\end{figure}

Atmospheric properties of KIC\,10920273 and KIC\,11395018 were
obtained from observations with the FIES spectrograph \citep{Frandsen1999} at the Nordic
Optical Telescope (NOT on La Palma, Spain), at medium resolution (R$\sim$46,000) in July and August 2010. The reduced spectra were
analyzed by several teams and the results were presented by
\citet{Creevey2012}. The constraints we used for our analysis are
shown in Table\,\ref{tbl_atm} (see also the 1- and 2-$\sigma$
error boxes in Fig.\,\ref{HR_AMP}). We adopted a set of atmospheric
constraints for each star that were closest to the
mean of results from several methods described by
\citet{Creevey2012}. This approach was preferred for the sake of
reproducibility, rather than using the mean values. However, we did
not restrict our model-searching space to less than 3-$\sigma$
uncertainty around these constraints; therefore, the selected values
represent well the overall results of the spectroscopic analysis.

The spectroscopic $\upsilon$\,sin\,$i$ values for KIC\,10920273 and KIC\,11395018 are 
$1.5\pm2.2$\,km\,s$^{-1}$ and $1.1\pm0.8$\,km\,s$^{-1}$, respectively \citep{Creevey2012}. These low values indicate either slow rotation or low inclination angle $i$, 
although the latter is statistically unlikely. Rotational periods from modulation of the $\textit{Kepler}$ data due to
spots on the surfaces of the two stars were measured to be $\sim$27 days for KIC 10920273 \citep{Campante11}, and $\sim$36 days for KIC 11395018 \citep{Mathur11}. The signal-to-noise ratios (SNR) of the peaks used for these measurements, particularly for KIC 10920273, were low, so the
results should be used with caution. \citet{Mathur11} inferred $i\geqslant45^{\circ}$ for KIC\,11395018 based on combining the rotational frequency splittings with the measured rotation period, which implies this star to be a slow rotator. A wider range of inclinations was possible for KIC\,10920273
\citep{Campante11}. However, the modulation in the light curve has been detected with less uncertainty in the
new analysis performed using longer time series including $\textit{Kepler}$ Q9
and Q10 data of KIC\,10920273 (Garc\'{i}a et
al., private communication). If confirmed, this would be consistent with a relatively high
inclination angle, and slow rotation. The effects of the
centrifugal force are negligible for slowly rotating stars.\footnote{If, counter to our expectations, one of these stars were to be confirmed as a fast rotator, rotational effects on the oscillation frequencies would have to be taken into account (see \citealt{Suarez2010} for a detailed analysis of the effects of centrifugal distortion on solar-like oscillations). Currently, there is no robust detection of rotational frequency splittings that can be included as constraints in our modeling.} However,
rotational mixing may lead to changes in the properties of the
models even for slowly rotating stars, because the efficiency of
this mixing is more directly related to differential rotation in
stellar interiors rather than to surface rotational velocities
\citep{Pinson1990,Eggenberger2010}. Studying the impact of
rotation on post-main-sequence stars would require a detailed
discussion of the effects of rotational mixing on the chemical
gradients in the central parts of the star. These influence the
asteroseismic properties of the models, particularly the mixed
modes. In the specific case of the evolved post-main-sequence stars
modeled here, however, we expect that these effects on the chemical
gradients would already be erased, as found by \citet{Miglio2007} for
models of 12 Bootis A in the thick-shell-H-burning phase (see also
the discussion of the effects of microscopic diffusion in the
subgiant HD 49385 by \citealt{Deheuvels2011}). We therefore have not included the rotational effects for most of the analyses (see Section\,3).

When the atmospheric properties alone are considered, these two
stars are very similar. Due to the degeneracy
inherent in the H-R diagram analysis (see, e.g., \citealt{Fernandes03}), it is not
possible to determine the global stellar properties with
sufficiently high precision to study their detailed
characteristics without the help of seismic data, which we now discuss.

\begin{deluxetable}{cccc}
\tablewidth{0pc}
\tablecolumns{4} \tablecaption{Adopted atmospheric constraints for KIC\,10920273 and KIC\,11395018 (from \citealt{Creevey2012}) \label{tbl_atm}}
\tablehead{\colhead{Star}& \colhead{$T_{\rm eff}$(K)} & \colhead{log\,\emph{g}} & \colhead{[Fe/H]} }
\startdata
KIC\,10920273& $5790\pm74$&$4.10\pm0.10$&$-0.04\pm0.10$\\
KIC\,11395018& $5700\pm100$&$4.10\pm0.20$&$0.13\pm0.10$\\
\enddata
\end{deluxetable}

\subsection{Asteroseismic data}

We used $\textit{Kepler}$ data from observations made in the period
from 2009 May to 2010 March, i.e., from the commissioning run (Q0) through
Quarter 4 (Q4). The formal frequency resolution is $\sim$0.05$\mu$Hz. From the power spectra, 
\citet{Campante11} and \citet{Mathur11} reported individual frequencies for KIC\,10920273 and KIC\,11395018, based on analyses performed by several teams. The final sets of
results included a minimal and a maximal list of frequencies, where the
former were those agreed upon by more than half of the fitters and the latter were those agreed upon by at least two fitters.
Therefore, the frequencies that are in the maximal list but not the
minimal list are less certain. For details of the
frequency-extraction techniques and the selection methods, we refer
the reader to \citet{Campante11} and \citet{Mathur11}. The analysis
of each star resulted in the extraction of up to a total of 25
individual oscillation frequencies for radial ($l=0$), dipole
($l=1$), and quadrupole ($l=2$) modes, including several mixed
modes. These mixed modes carry information
from the core and hence provide stronger constraints on the
evolutionary stage of the stars, as discussed in Section~1. We
started by searching for models using the minimal-list of frequencies
and then extended our analysis to include additional frequencies from the
maximal lists.

\section{Modeling approach}

Asteroseismic modeling is performed by optimizing the stellar model
parameters to match the observed seismic quantities and also the classically observed (or derived) stellar properties, such as effective temperature, surface gravity,  surface metallicity (along
with radius, mass and luminosity, when available). The seismic
quantities include, but are not limited to, the average large and
small frequency separations\footnote{The large frequency separation is $\Delta\nu_{n,l}=\nu_{n,l}-\nu_{n-1,l}$, and the small frequency separation is $\delta\nu_{n,l}=\nu_{n,l}-\nu_{n-1,l+2}$, where $\nu_{n,l}$ is the frequency of the mode with spherical degree $l$ and radial order $n$.}, the frequency of maximum power in the oscillation spectrum ($\nu_{\rm max}$), and the
individual oscillation frequencies. Naturally, the individual
frequencies provide the most detailed information and the highest
precision in the derived stellar properties (see, e.g.,
\citealt{metcalfe10ApJ} and \citealt{mathur2012}).

We used the individual oscillation frequencies and the atmospheric
properties ($T_{\rm eff}$, $\log g$, and [Fe/H]) as constraints to
carry out the stellar model optimization. As an initial guess for
the parameter space to be searched, we used preliminary results of the mass determination from the analysis of \citet{Creevey2012}, which were in agreement with their final results within the uncertainties. They derived stellar properties using the average seismic quantities together with the atmospheric constraints. The final values given by \citet{Creevey2012} were $1.25\pm0.13~M_{\odot}$ for KIC\,10920273 and
$1.37\pm0.11~M_{\odot}$ for KIC\,11395018.

Five teams participated in the modeling of these stars using a variety of
evolutionary codes and fitting methods. Most of the methods were either based on searching for the best-fitting model in
a grid specifically computed for this analysis or on using a pre-existing grid to determine the general area of the stellar properties in the parameter space before going into further refinement process for individual stars. One team used the Asteroseismic Modeling Portal
(AMP), which is a pipeline analysis tool that optimizes the seismic
and non-seismic properties globally using a genetic algorithm. AMP starts the model-search with four random independent sets of initial parameters
and performs the search over a large parameter space \citep{metcalfe09,
woitaszek2009}. The variety of codes and methods employed give us an
estimate of the external uncertainties inherent in the analysis. The
list of codes and the configurations regarding the input physics are
presented in Table~\ref{input}.

The individual fitting methods also differed slightly. ASTEC1
calculated grids of models within the 3-$\sigma$ uncertainty of the
non-seismic constraints and performed the optimization by a 2-step
process, refining the grids several times in the second step guided by the seismic
$\chi^2$ values -- described by Equation~(\ref{chi2}). ASTEC2
explored the models, which included turbulent diffusion, and calculated
individual models guided by the frequencies. CESAM looked for models
reproducing the first avoided crossing as an initial requirement and
then performed an optimization using $\chi^2$-minimization to
determine stellar mass and age (see \citealt{Deheuvels2011} for
details of this method). The Geneva stellar evolution code was used
to compute grids of rotating models with an initial velocity of
50\,km\,s$^{-1}$ on the zero-age main sequence (ZAMS). This value
results in surface velocities that are typically lower than
10\,km\,s$^{-1}$ at the end of the main sequence (MS) for a
solar-type star that is assumed to undergo magnetic breaking on the
MS due to the presence of a convective envelope
\citep{Krishnamurthi1997}. The initial parameters used by each team
are given in Table~\ref{grids}.

Oscillation frequencies of low-degree modes were calculated by
LOSC \citep{Scuflaire2008} for stellar models computed by CESAM,
while the Aarhus Adiabatic Pulsation Package (ADIPLS,
\citealt{cd08adipls}) was used to calculate the frequencies for all
of the other models.

In relation to the oscillation frequencies, there is a well-known offset between the observed and the
model frequencies for the Sun and solar-type stars, due to inaccurate representation of the
near-surface layers in the models. To address this
issue, we used the empirical correction suggested by
\citet{kjeldsen08}, who showed that the difference between the
observed and calculated solar frequencies, which gets larger with increasing frequency, can be fitted by a power law:

\begin{equation}
\nu_{\rm obs}(n,0)-\nu_{\rm best}(n,0)=a\left( \frac {\nu_{\rm
obs}(n,0)}{\nu_0}\right)^b, \label{correction}
\end{equation}
where $\nu_{\rm obs}(n,0)$ and $\nu_{\rm best}(n,0)$ are the
observed and best model frequencies with degree $l=0$ and radial
order $n$, $\nu_0$ is a constant frequency usually chosen to be the
frequency at maximum oscillation power, $a$ is the size of the correction at
$\nu_0$ and can be calculated for each model, and $b$ is the exponent to be determined.

The right-hand-side of Equation~(\ref{correction}) is the
correction term to be added to the acoustic (p-mode) frequencies of
the best models. The mixed modes, however, are less sensitive to the
properties of the near-surface layers since much of their energy is
confined to the stellar center. In other words, we need to
apply a smaller near-surface correction to the frequency of a mixed
mode than to a p mode with a similar frequency. Following \citet{Brandao2011}, we scaled the magnitude of the
correction inversely with $Q_{nl}$, the inertia of a given
mode normalized by the inertia of a radial ($l=0$) mode at the same
frequency (see, e.g., \citealt{aerts10}). Note that the inertia of a
mixed mode is much higher than that of a p mode. The correction to
be applied to all calculated frequencies is then of the form:
\begin{equation}
\nu_{\rm corr}(n,l)=\nu_{\rm
best}(n,l)+a\left(\frac{1}{Q_{nl}}\right)\left( \frac {\nu_{\rm
best}(n,l)}{\nu_0}\right)^b, \label{corr_Qnl}
\end{equation}
where $\nu_{\rm corr}$ represents the corrected model
frequencies. Notice that $\nu_{\rm obs}(n,0)$ on the right-hand-side
of Equation~(\ref{correction}) is replaced by the best model
frequencies in order to allow us to correct the frequencies outside
the range of observed radial modes (see,
\citealt{Brandao2011}, for details).

The solar value of the exponent $b$ was calculated by
\citet{kjeldsen08} to be 4.90 using the GOLF data \citep{lazrek97} and
the solar ``Model S'' of \citet{modelS96}. This value was found to range
from 4.40 to 5.25 for the same model, depending on the number of
radial orders included in the calibration, but $a$ was
found to vary by less than 0.1\,$\mu$Hz in all cases.
\citet{kjeldsen08} suggested that the solar $b$ value may be used
for solar-like stars and this approach was successfully applied to $\beta$~Hyi \citep{Brandao2011}, HD\,49385 \citep{Deheuvels2011}, KIC\,11026764
\citep{metcalfe10ApJ} and a sample of
\textit{Kepler} stars \citep{mathur2012}. In this work ASTEC1, ASTEC2, and Geneva codes adopted the solar value $b=4.90$
from \citet{kjeldsen08} for calculating the correction term, while
AMP adopted $b=4.82$, which is the solar-calibrated value for AMP
with the BiSON data \citep{chaplin99}, and CESAM adopted $b=4.25$,
the calibrated value using the GOLF data \citep{Gelly2002}. Given
how little $a$ varies for a relatively large range of $b$ for a given model, as discussed above, using slightly different $b$ values for model-fitting has a negligible impact on the results.

%!!!MOVED TWO TABLES FROM HERE TO THE END after References

\section{Results and Discussion}
\subsection{Global properties}

To select the best models, we defined the two normalized $\chi^2$
measures shown in Equations~(\ref{chi2})~and~(\ref{chi2_atm}), which
allowed us to evaluate the qualities of the fits for the atmospheric
parameters and the seismic parameters separately. The seismic
measure was

\begin{equation}
\chi^2_{\rm
seis}=\frac{1}{N}\sum_{n,l}\left(\frac{\nu_{\mathrm{obs}}(n,l)-\nu_{\mathrm{corr}}(n,l)}{\sigma(\nu_{\mathrm{obs}}(n,l))}\right)^2,
\label{chi2}
\end{equation}
where $N$ is the number of observed frequencies,
$\nu_{\mathrm{corr}}(n,l)$ represents the near-surface-corrected
model frequencies with spherical degree $l$ and radial order $n$,
$\nu_{\mathrm{obs}}(n,l)$ are the observed frequencies, and
$\sigma(\nu_{\mathrm{obs}}(n,l))$ are the uncertainties on the observed
frequencies. The measure for the atmospheric properties was

\begin{equation}
\chi^2_{\rm atm}=\frac{1}{3}\sum
\left(\frac{\rm P_{\mathrm{obs}}-\rm P_{\mathrm{mod}}}{\sigma(\rm P_{\mathrm{obs}})}\right)^2,
\label{chi2_atm}
\end{equation}
where P=\{$T_{\rm eff}$, log\,\emph{g}, [Fe/H]\} and the subscripts
``obs" and ``mod" represent the observed and model properties,
respectively, with $\sigma(\rm P_{\mathrm{obs}})$ denoting the
observational uncertainties. The values of [Fe/H] for the
models were calculated using the formula
${\rm[Fe/H]}=\log(Z/X)_{\rm mod}-\log(Z/X)_{\odot}$, where the solar
value was adopted from \citet{grevesse93} as
$(Z/X)_{\odot}=0.0245$.

Each modeling team returned the model that best matched the
observational constraints. We present the properties of these models
in Tables~\ref{scully} and \ref{boogie}, along with the normalized
$\chi^2$ values\footnote{The labels of the models presented in
Tables 4 and 5 start with ``S"  and ``B" for KIC\,10920273
and KIC\,11395018, which stand for ``Scully"  and ``Boogie" -- the
nicknames of the stars within Kepler Asteroseismic Science
Consortium, Working Group 1.}. We also present models fitted using
more or fewer frequencies than those in the minimal lists in order
to see whether the model-fitting results change considerably.
In each case, we calculated the $\chi^2_{\rm seis}$ in the tables using
only the frequencies that were common constraints for all of the
models, in order to achieve a consistent evaluation of the
models.

\renewcommand{\labelitemi}{\textgreater}

\begin{figure}[h]
\begin{center}
$\begin{array}{c}
\includegraphics[scale=0.36,angle=90]{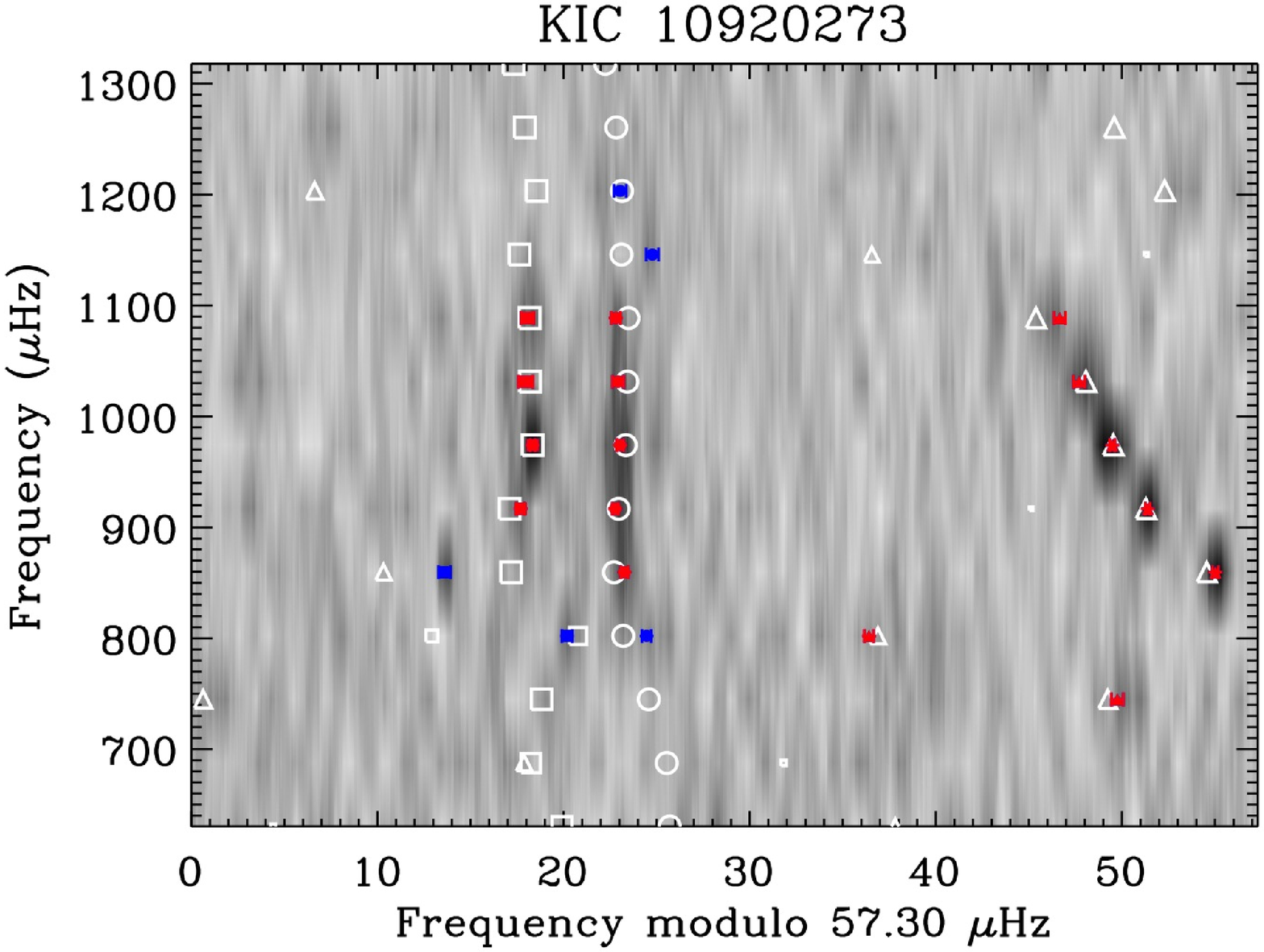}\\
\includegraphics[scale=0.36,angle=90]{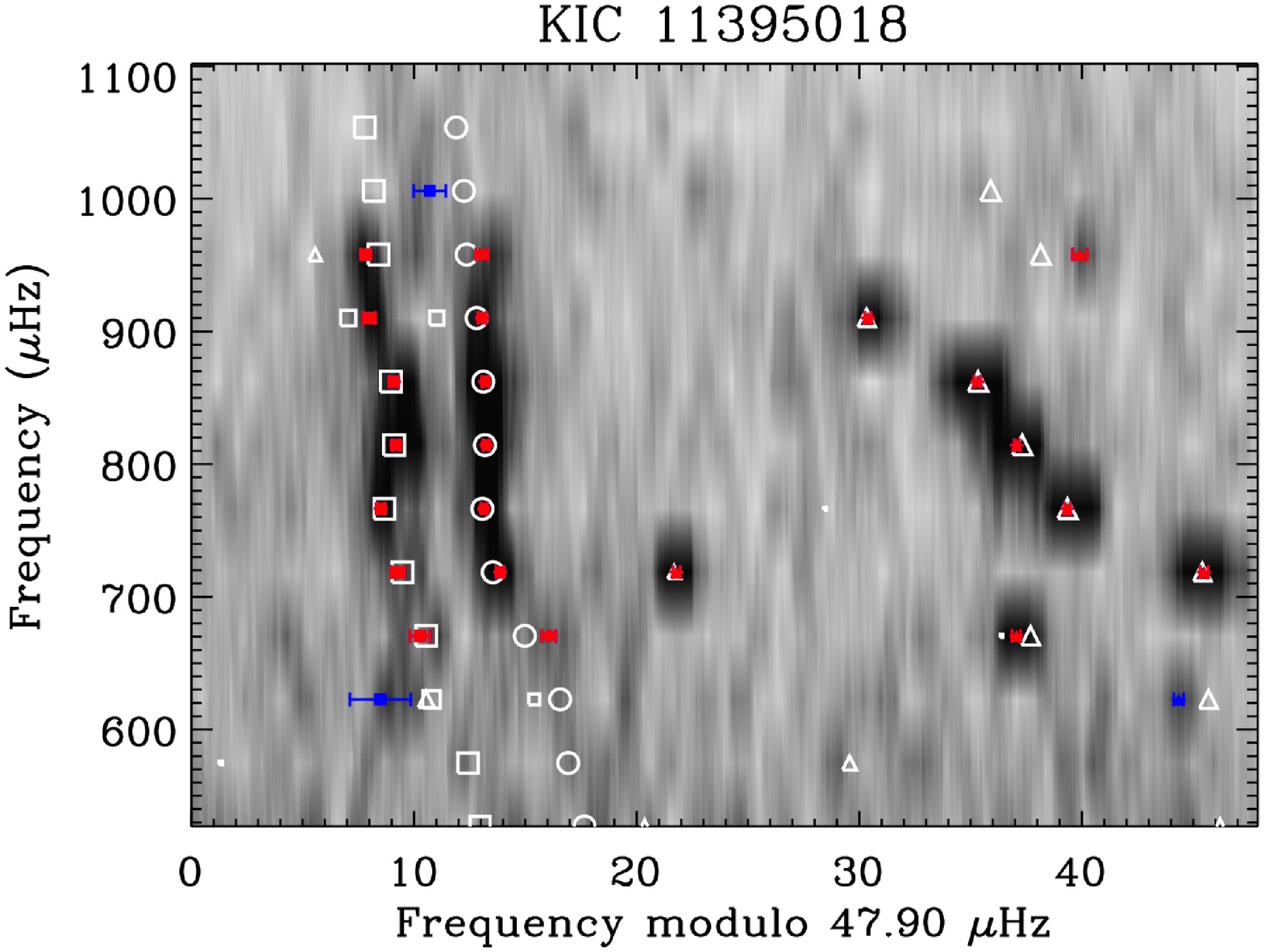}\\
\end{array}$
\end{center}
\caption{\footnotesize \'{E}chelle diagrams for the selected models
of KIC\,10920273 (SA3, upper panel) and KIC\,11395018 (BD, lower
panel). The radial orders
are plotted horizontally (see \citealt{Bedding2012}) and the vertical axes indicate the
frequencies at the middle of each order. The background is a gray-scale map of the power spectrum
from the observations, with the observed frequencies shown by
filled symbols with 1-$\sigma$ error bars. Minimal-list frequencies
are plotted in red, while the additional maximal-list frequencies
are plotted in blue. Open symbols are used for the model
frequencies, with smaller size implying larger normalized mode inertia. Circles, triangles, and squares represent $l=0$, $l=1$,
and $l=2$ modes respectively. } \label{boogie_AMP}
\end{figure}

There is a good agreement between the observed
frequencies and the model frequencies. The quality of the fits can be seen in the  \'{e}chelle diagrams for a sample of models (Fig. \ref{boogie_AMP}). Overall frequency patterns, including the dipolar mixed modes, are matched quite well. 
The fact that $\chi^2_{\rm seis} \textgreater 1.0$ implies that
either the observational uncertainties are underestimated or the models are incomplete representations of the observational data. The models with
relatively high $\chi^2_{\rm seis}$ ($\gtrsim20.0$), are those that either could be improved with further refinement or that do not reproduce
all of the modes simultaneously, in particular the mixed modes, which are more difficult to fit. Due to
their strong sensitivity to stellar evolution, mixed modes tend to
dominate the model-fitting results. The timescale on which the
signatures of these modes evolve is very short, so it becomes difficult
to find good fits unless the grid of models used is very fine.

We present our results in Tables~\ref{scully} and \ref{boogie}.  Properties of all the models are listed, along with the weighted mean values and the standard deviations. Both stars have left the main sequence (central hydrogen mass fraction $X_{\rm c}=0.0$) but
have quite different characteristics, as seen in Tables~\ref{scully} and \ref{boogie}. The fact that KIC\,10920273 is an old solar analog (with one solar mass and near-solar metallicity) makes it  an interesting target for further studies. A typical solar model would turn off from the MS at around 9-10 Gyr. However, the metallicity and particularly the helium abundance alter this age estimate. In this case it is the high helium abundance that affects the MS turn-off age more than the low metallicity. The models with higher helium abundance behave similar to those with higher mass (higher luminosity), following an evolutionary track similar to that of a higher-mass star, and hence have shorter MS lifetimes.

The results in Tables~\ref{scully} and \ref{boogie} were weighted by the goodness of the seismic fit, i.e. the inverse of $\chi^{2}_{\rm seis}$. This way, any misleading contribution due to coarse grids is
eliminated. Although this does not provide a direct measurement of
the systematic uncertainties, it still allows us to estimate the
order of magnitude of the external errors expected from using
different inputs, codes, and fitting methods. A similar determination of the systematic errors for the case of bright $\textit{Kepler}$ stars 16 Cyg A and B  was carried out by \citet{metcalfe2012} using different evolutionary codes and fitting methods. The uncertainties we determined are mostly greater due to the lower SNR in the data of our faint stars. Systematic errors in determination of stellar properties using grid-based pipelines caused by different observational constraints and different input physics were discussed more generally for a few \textit{Kepler} stars by \citet{Creevey2012}. We discuss the uncertainties further in the next section.

%MOVED TWO TABLES FROM HERE TO THE END AFTER the OTHER TWO TABLES

\subsection{Discussion of uncertainties}

To evaluate the typical uncertainties caused by different input physics further, we calculated some additional models and small grids using KIC\,10920273 as a test case. We selected SB1 as our base model.  Keeping the input parameters (mass, $Z/X$, $Y$, $\alpha$) fixed, we first explored the effects of changing one single ingredient of input physics at a time. We also changed the value of $\alpha$ while keeping everything else fixed. On every new evolutionary sequence, we calculated the frequencies of the models that had all atmospheric properties ($\log g$, $T_{\rm eff}$, and [Fe/H]) within 3\,$\sigma$ of the observed values. We then selected the model that best matched the observed frequencies. In the cases of including core overshoot, changing the convection treatment (to CGM formulation), and using several different nuclear rates \citep{bahcall95,parker86, adelberger98}, the new models reproduced the observed frequencies as well as the base model with an age difference of only 0.7$\%$ at most (which corresponds to changes in radius of $\textless 0.2\%$ and in $\log g$ of $\textless 0.1\%$). This means that the uncertainties on the final parameters caused by the corresponding inputs were negligible. We note that the effect of core overshoot would be more significant for a main-sequence star.

For the cases where we did not obtain a model having a seismic $\chi^2$ comparable to that of the base model, we carried on with the analysis. These cases resulted from including diffusion and gravitational settling of helium, using two different versions of the low-temperature opacities given in Table~\ref{input}, and varying the value of mixing length parameter (in the range of 1.6--2.0). For each of these cases, we computed additional small grids around the base model by varying all the input parameters, in order to see how much the output properties were different for two models with different input physics but similar frequencies. We then calculated the weighted mean and standard deviation in the same way as in the original analysis. The standard deviation in this case represents the typical uncertainties caused by using a fixed set of input physics, hence decreasing the level of model-dependance in the results substantially. The mean values for age, luminosity, radius, $T_{\rm eff}$, and $\log g$ calculated from the additional analysis agreed with the original results within 1\,$\sigma$, while their standard deviations were of the same order as those qiven in Table \ref{scully}. This confirms that the uncertainties presented here are realistic. Moreover, the resulting values of radius and $\log g$ from the additional analysis are essentially the same as the original results. This is reassuring given the importance of asteroseismology in determining the radius in a robust way.

The internal uncertainties were different for each method. However,
the dominant source of uncertainty is the non-uniqueness of the
solution rather than the statistical errors. Parameter correlations allow
a trade-off between parameters, leading to different families of solutions
that are almost equally good seismic fits. The effective range of these
correlations is narrowed substantially, but not eliminated, by the use
of seismic data. We used a single method (AMP) to evaluate the uniqueness of the
best models for both stars, since the uniqueness depends on the specific
constraints adopted in each case. AMP finds the lowest value of $\chi^2$
in the entire search space. Along the way it also identifies the secondary
minima -- which can be far away from, but not much worse than, the
formally best solution. For both stars,
these secondary minima (SA2 and BA2 in Tables~\ref{scully}~and~\ref{boogie}) are marginally worse than the best solutions found using the same set of inputs (SA1 and BA1), but
the values of the mass and helium abundance are quite different. This reflects the well-known mass-He degeneracy (see, e.g., \citealt{lebreton93, Fernandes03, metcalfe09}), which is not
entirely lifted, even with the help of asteroseismology. The mass is
constrained more strongly than it would be without asteroseismic data but in the absence of external constraints on the helium abundance, we cannot
choose one particular model. Consequently, we used both the primary- and
secondary-minimum solutions from AMP in the calculation of the weighted
means. The range of the AMP results leads to relatively large ``uniqueness
uncertainties'' in the fitted stellar properties. These are determined as follows for KIC\,10920273, and KIC\,11395018, respectively: 3\%, and 7\% in stellar
mass; 8\%, and 21\% in the initial metallicity $(Z/X)_{\rm i}$; 10\%, and 25\%
in $Y_{\rm i}$; 7\%, and 8\% in age; and 1\%, and 2\% in radius. Note that the larger uncertainties in the mass and
chemical composition for KIC\,11395018 reflect the fact that it has
\textit{two} observed avoided crossings, which provide more stringent
constraints such that neighboring models are significantly worse. Only a
relatively large jump along the parameter correlations yielded a
secondary minimum with a comparable seismic fit \citep[cf.][]{metcalfe10ApJ}.
When we evaluate the uncorrelated statistical uncertainties for each of
these minima with a local analysis using singular value decomposition
(SVD), we found very small errors (e.g., as low as 5$\times$10$^{-3}\%$ for mass), reflecting the steep $\chi^{2}$ surfaces corresponding to these results. Although these 'local' uncertainties are real, i.e., changing the values of the parameters by the correlated 'tiny' uncertainties causes a large increase in the $\chi^{2}$ ($\textgreater50$), the minima are not unique. Hence, a
local analysis does not reflect the true uncertainties of the overall results.

Inclusion of the maximal list of frequencies as constraints in the
analysis (five additional frequencies for KIC\,10920273 and two for
KIC\,11395018) made the agreement between the model and observations
better for  both KIC\,10920273 (SA3 in Table~\ref{scully}) and
KIC\,11395018 (see BA3 in Table~\ref{boogie}). We also used a third
list for KIC\,10920273 to ensure that the less-certain frequencies did
not bias our model-fitting. For that purpose, we left out two frequencies in the analysis. One of these was identified as a dipole-mode
frequency belonging to the minimal list, but tagged as being close
to the second harmonic of the long-cadence frequency
(with $\nu=1135.36\pm0.31~\mu$Hz), while the other one was identified as a quadrupole
mode (with $\nu=873.10\pm0.32~\mu$Hz) that was tagged as a possible
mixed mode introduced a posteriori \citep{Campante11}. Both SA4 and SB2 were selected using
this alternative frequency set for KIC\,10920273, and the agreement
between the model and observations was not affected substantially. Therefore, we
cannot ascertain whether these two peaks are stellar in origin. Nevertheless, we do not completely rule out the possibility of these peaks being stellar as some of the models that result from using alternative frequency sets do contribute to the weighted mean values significantly. Furthermore, we note that the peak at $\nu=873.10~\mu$Hz, which is tagged as a possible quadrupole mixed mode, is in the middle of the frequencies of a dipole mixed mode and a quadrupole mode in most of our models. So the observed peak may correspond to a dipole mixed mode, which, according to the models, has a relatively low normalized mode inertia indicating an observable amplitude.

%\newpage
\subsection{Comparison with previous results}

Comparing our results with those from the pipeline analyses of Creevey et al.
(also given here in Tables~\ref{scully} and \ref{boogie}), we see
that the mass determinations from the pipeline analyses were higher,
which led to lower age estimates. We emphasize that the previous
pipeline analyses used only the average seismic quantities, hence
lacking additional information from the individual frequencies and
being affected by the uncertainties of the scaling relations.
Therefore, some deviation from their values was expected.
Nonetheless, it is reassuring that the mass, radius, and age determinations of \citet{Creevey2012} are within 2-$\sigma$ uncertainty limits of our results for
KIC\,10920273, and within 1\,$\sigma$ for KIC\,11395018. We also confirm the robust
determination of $\log g$ using scaling relations and grid-based
analyses (see Table\,7 in \citealt{Creevey2012}), with which our results are
in agreement within 1\,$\sigma$. Additionally, we note that our results confirm that the mass and radius determined using only the scaling relations (e.g., \citealt{mathur2012}) provide good initial estimates for these properties ($1.06\pm0.20M_{\odot}$ and $1.80\pm0.11R_{\odot}$ for KIC~10920273; $1.31\pm0.25M_{\odot}$ and $2.21\pm0.14R_{\odot}$ for KIC~11395018).

There is excellent agreement, for both stars, between our results and the mass
estimates of \citet{Benomar2012}, who used the coupling strength of
the observed mixed modes to determine the masses of several subgiants, including KIC~10920273 ($1.04 \pm 0.04 (\pm 0.04) M_{\odot}$) and KIC~11395018 ($1.21 \pm 0.06 (\pm 0.04) M_{\odot}$).

The most substantial improvement in this work comes from the use of
individual frequencies which yield increased precision, with age
being affected the most. The presence of the mixed modes in the data
allowed us to determine the age with 5-7\% precision, although with some model-dependency. This result is a 
major improvement on the 35-40\% precision in age achieved using atmospheric and mean seismic parameters. Both stars are determined to be post-main-sequence
subgiants with no hydrogen left in their cores. Evolutionary tracks
of the selected models are shown in Fig.\,\ref{HR_AMP}.

Although we did not restrict the parameter search to be within
1-$\sigma$ uncertainty around the spectroscopic constraints, the
weighted mean values of $T_{\rm eff}$ from the models are within
1-$\sigma$ limit for both stars, while $\log g$ results are in
agreement with the spectroscopic values within 2\,$\sigma$ (see Fig.\,\ref{HR_AMP}), and [Fe/H] within 1.5\,$\sigma$. Our $\log g$ results are in excellent agreement with asteroseismic $\log g$ values obtained from scaling relations (given by \citet{Creevey2012} and also in Tables\,\ref{scully} and \ref{boogie}). We
also note that our temperature results are in good agreement with
the revised photometric values for the Kepler Input Catalog (KIC)
from \citet{Pinson2011}, who derived $T_{\rm eff}=5872\pm70$\,K for
KIC\,10920273, and  $T_{\rm eff}=5650\pm59$\,K for KIC\,11395018.

\subsection{Non-seismic age diagnostics}

We discussed the asteroseismic constraints on the stellar age in Section\,1. Here we discuss the implications of rotation and stellar activity on the age, as well as those of the surface lithium abundance.

Rotation and activity are potentially valuable diagnostics of stellar age. It was shown that the Ca$^{+}$ emission luminosity, an indicator of stellar activity, decays roughly as $t^{-0.5}$ for some cluster stars and the Sun \citep{Skumanich72}, furthermore, rotational decay was shown to follow  the same law. Large samples of stellar rotation periods have been collected, and the $\textit{Kepler}$ mission promises many more.  There is therefore substantial interest in stellar rotation-mass-age, or gyrochronology, relations (see \citealt{Barnes2003,Barnes2007,Mamajek2008,Meibom2011,Epstein2012}). We discussed in Section\,2.1 that relatively slow rotation is inferred for both stars.  Slow rotation rates imply relatively old stars, which is consistent with our asteroseismic determinations. However, one would not expect main-sequence spin-down relationships to apply directly to the evolved stars. Therefore, we cannot use the rotation rates for these stars to infer their ages with the age-rotation relations established for MS stars; these relations need to be calibrated for more evolved stars using larger samples.

\begin{figure*}[!ht]
\begin{center}
$\begin{array}{cc}
\hspace{-0pt}\includegraphics[scale=0.45,angle=0]{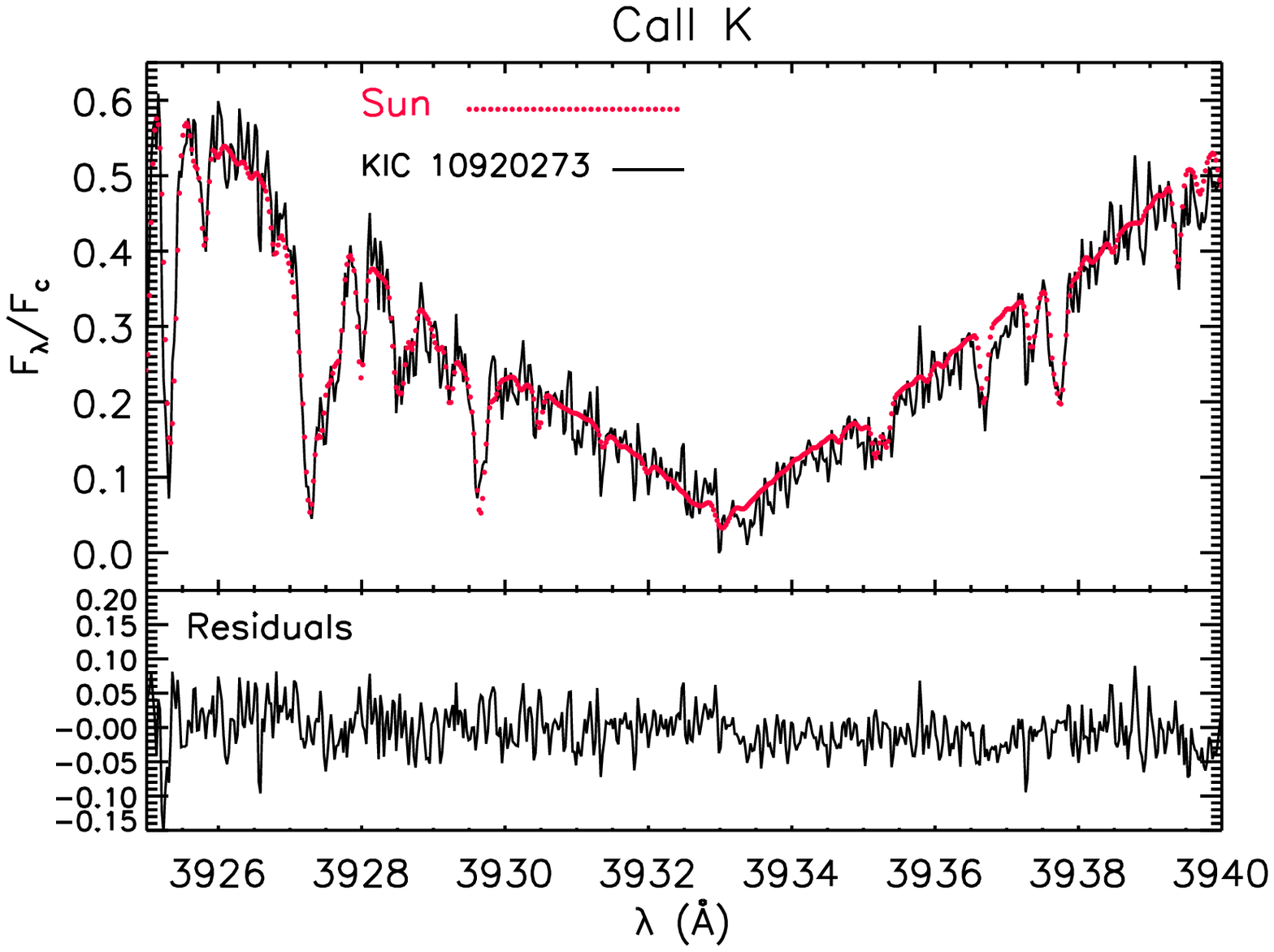} &\hspace{-0pt}\includegraphics[scale=0.45,angle=0]{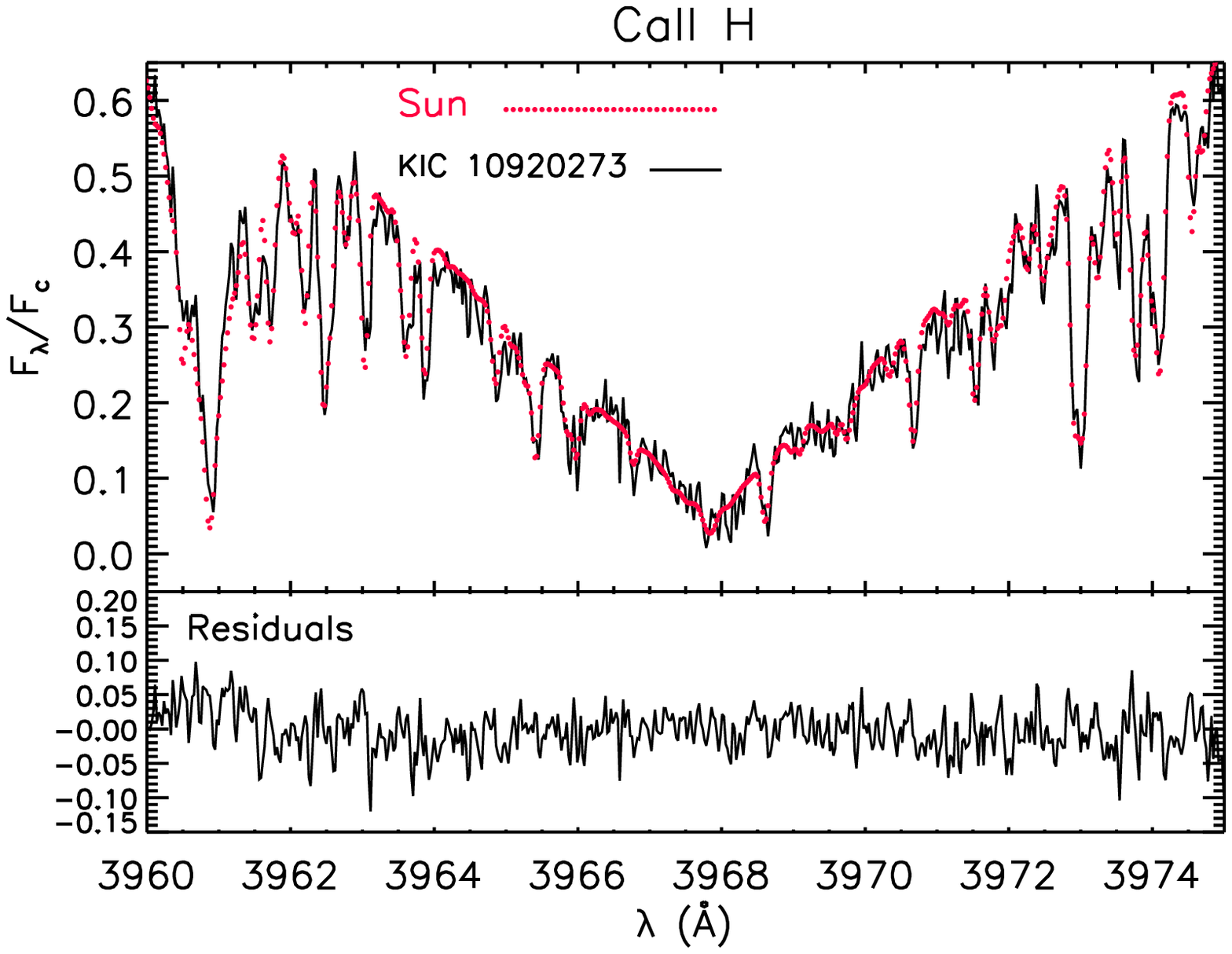}\\
\hspace{-0pt}\includegraphics[scale=0.45,angle=0]{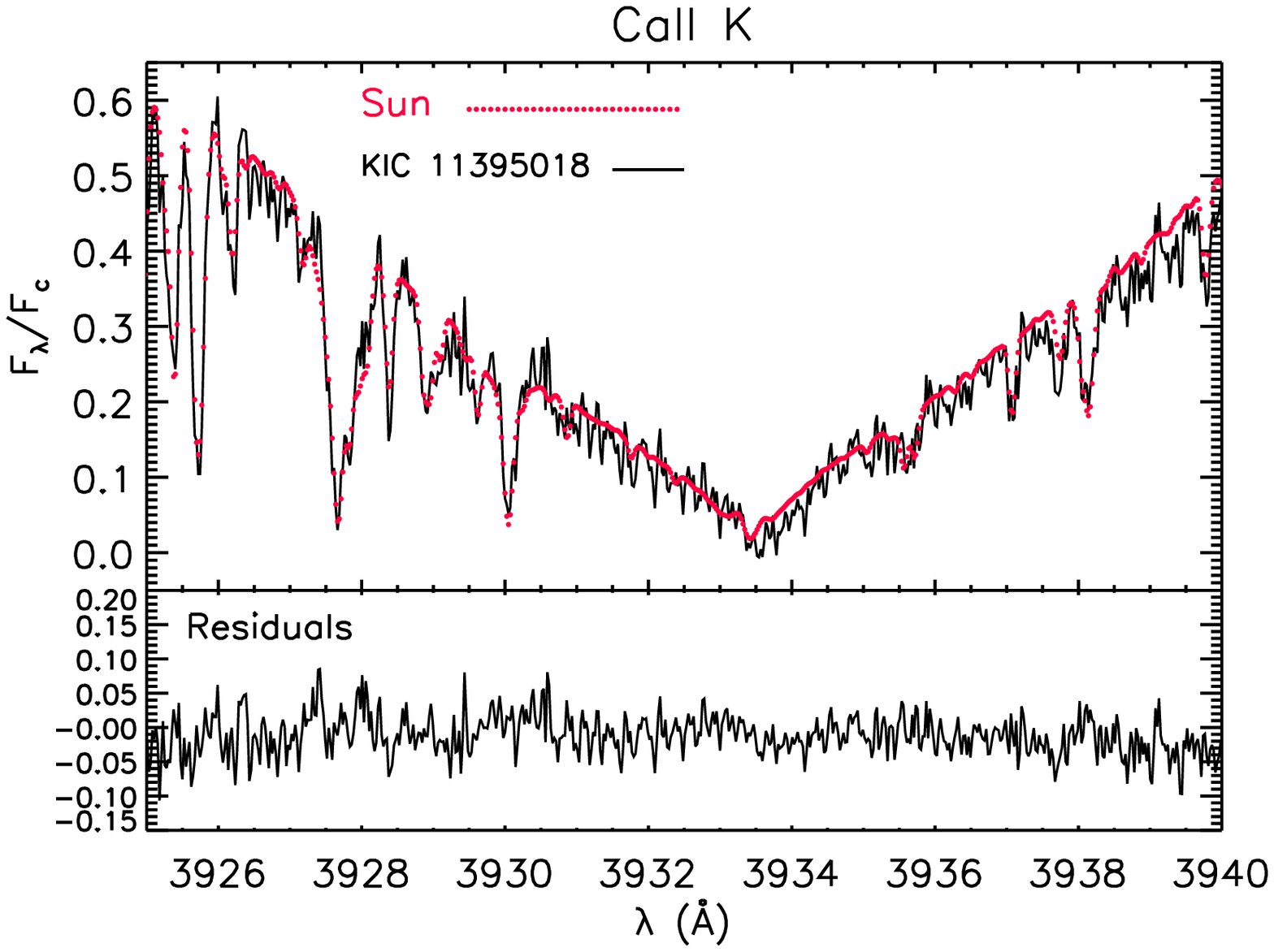}
&\hspace{-0pt}
\includegraphics[scale=0.45,angle=0]{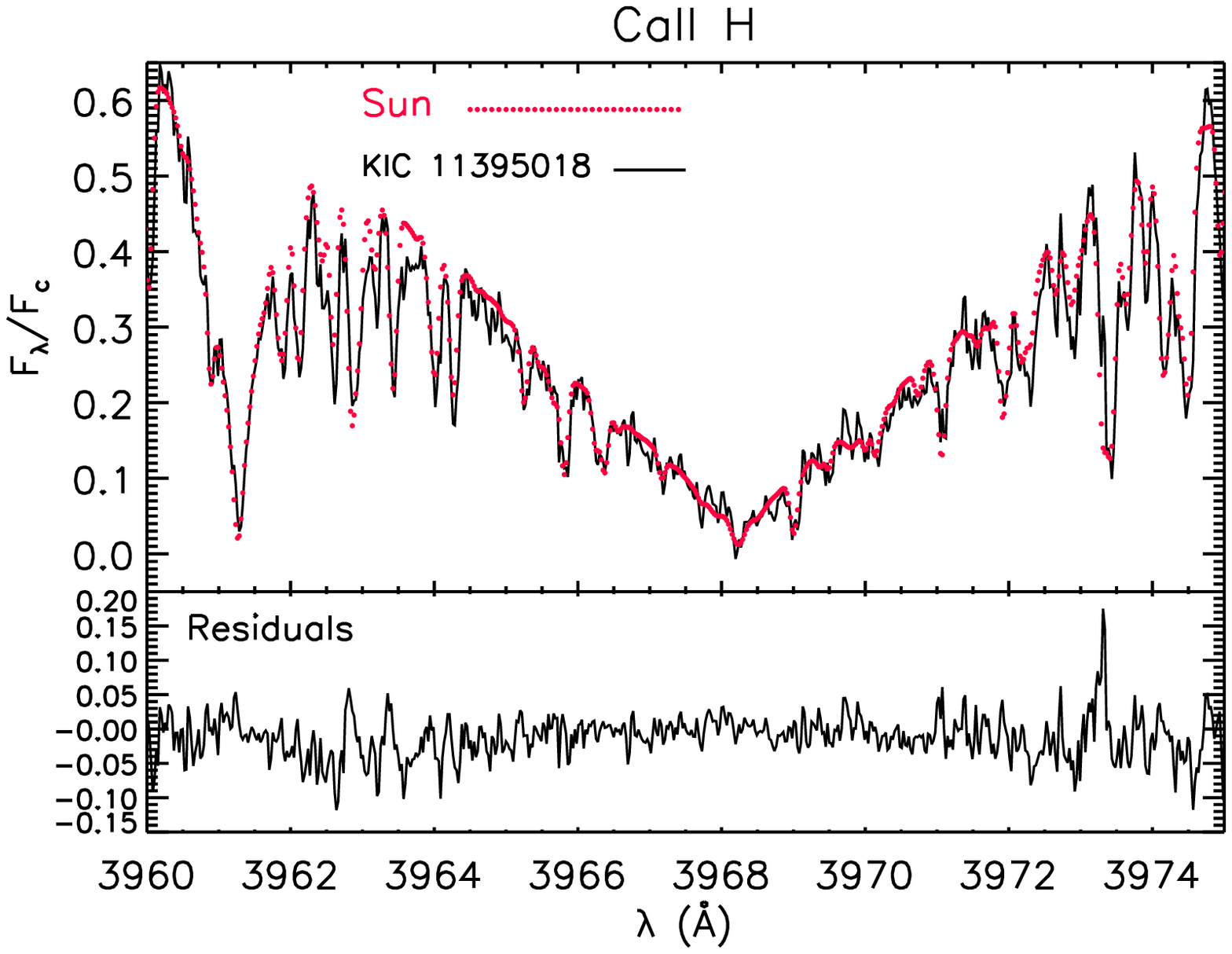}\\
\end{array}$
\end{center}
\caption{\footnotesize Chromospheric activity in the Ca{\sc
ii}~K and Ca{\sc ii}~H lines (flux relative to the continuum) for KIC\,10920273 (upper panels) and
KIC\,11395018 (lower panels). The solar spectrum (Ganymede taken in
2007 with HARPS) is overplotted with a dotted (red) line. The
residuals between the stellar and solar spectra (at the bottom of
each plot) show that the two stars have activity levels comparable
to the quiet Sun or lower.} \label{activity}
\end{figure*}

We have analyzed the chromospheric activity in the Ca{\sc
ii}~HK lines and found the levels of activity in
both stars to be very low. Fig.~\ref{activity} shows the Ca{\sc ii}~K
and H lines of KIC\,10920273 and KIC\,11395018 compared to the Sun.
The solar spectrum was obtained from the solar light reflected by
Ganymede, which was observed with HARPS in April 2007\footnote{\tt
http://www.eso.org/sci/facilities/lasilla/instruments/ harps/inst/monitoring/sun.html},
when the Sun was close to the minimum of its activity cycle. We
accounted for the different resolving power of HARPS ($\rm R\simeq
120,000$) compared to FIES spectrograph ($\rm R\simeq 46,000$) by
convolving the solar spectrum with a Gaussian kernel of the appropriate width. It is clear that these two \textit{Kepler} stars
have chromospheric activity levels comparable to the quiet Sun, or
lower. These low activity levels are consistent with the old ages we infer from asteroseismology; however, the rough nature of the empirical
age-activity relations for post-MS stars does not allow us to make a quantitative analysis to infer ages. 

Another independent determination of stellar age may be obtained by measuring the Li content at the stellar surface. Lithium is
easily destroyed in stellar interiors and is only produced under
unusual circumstances; it has therefore been employed as an age
indicator for low-mass stars. Lithium can be directly depleted if
the surface convection zone is deep enough. It can also be mixed into the
radiative interior, or it can be stored below the surface convection
zone by microscopic diffusion processes. In standard stellar models,
pre-main-sequence depletion occurs for most low-mass stars when they
have deep convection zones, and it is most severe in lower mass
stars \citep{Bodenheimer1965}. In a qualitative sense, a detection
of Li in very cool stars is a strong indicator of youth.  However,
standard models also predict that stars of order 0.9 solar
masses and higher would not experience main-sequence Li depletion,
and there is strong evidence from open clusters for a steady
decrease in Li as a function of time, even for stars more massive
than the Sun (see \citealt{Zappala1972}; \citealt{Pinsonneault1997};
and \citealt{SestitoRandich05} for reviews.)

It was also shown by  \citet{Randich2010} that, for a fraction of
solar-like stars with effective temperatures between 5750~K and
6050~K, Li is not further depleted after the age of $\sim$1\,Gyr, unlike
for the Sun and many other stars. Due to this bimodal pattern, Li
abundance alone cannot be used to determine the age for all
stars, and a high Li abundance can only help define a lower limit
for the age, since it may correspond either to a young star that has not yet
depleted much Li, or to an older star that has stopped depleting Li a
long time ago.

\citet{Creevey2012} showed that the two stars considered here have
strong Li absorption lines, which implies a high Li content at the
surface (log $N$(Li)$=2.4\pm0.1$ for KIC\,10920273 and log
$N$(Li)$=2.6\pm0.1$ for KIC\,11395018; where $\log N({\rm Li})=\log \,[n({\rm Li})/n({\rm H})]+12$, with $n$ being the number density of atoms and $\log N({\rm H})=12$ by definition).  Considering the empirical Li-age relation established by
\citet{SestitoRandich05}, Creevey et al. then determined that the
given Li abundances would indicate low ages (1-3 Gyr for
KIC\,10920273 and 0.1-0.4 Gyr for KIC\,11395018), which are
incompatible with the asteroseismic ages they determined through the
pipeline modeling (see Tables~\ref{scully} and \ref{boogie}) performed using the average asteroseismic quantities.

However, in addition to the bi-modality mentioned above, the age-Li relation has been shown to be valid for MS stars and does not necessarily extend to more evolved stars. This makes the age determination using the Li abundance
ambiguous. Thus, despite the high Li abundance, we are confident that these are indeed evolved stars that have left the main sequence, due to the presence of the mixed modes
in the observed oscillation spectra and as confirmed by our asteroseismic analysis.

\section{Summary and Conclusions}

We performed asteroseismic modeling of two $\textit{Kepler}$ stars,
KIC\,10920273 and KIC\,11395018, for which we have long seismic data sets ($\textgreater 8$ months) and ground-based follow-up spectra. We used
individual oscillation frequencies and atmospheric properties
as initial constraints. We employed several evolutionary codes with
different input physics, and various fitting methods to determine the global stellar properties and estimate
their uncertainties (see Tables \ref{scully} and \ref{boogie}). The
near-surface correction was applied to the models, which reproduced
the individual observed frequencies with considerable success; see
Fig.\,\ref{boogie_AMP} for a qualitative representation with
\'{e}chelle diagrams. These two relatively faint stars, which have similar atmospheric properties
according to the ground-based data, turned out to be substantially
different -- more than could have been predicted from their
different metallicities -- after incorporating the high-precision
asteroseismic data into the modeling.

KIC\,10920273 resembles an old Sun, having one solar mass
($1.00\pm0.04M_{\odot}$) and an age of $\tau=7.12\pm0.47$\,Gyr,
while KIC\,11395018 has a mass of $1.27\pm0.04M_{\odot}$ and an
age very close to that of the Sun ($\tau=4.57\pm0.23$ Gyr). These results agree, at the 2-$\sigma$ level for KIC\,10920273 and
1-$\sigma$ level for KIC\,11395018, with the properties determined
using the average asteroseismic quantities. The results
presented here are much more precise than those from the average asteroseismic quantities, though, and they are also more
accurate as a result of using more observational information, i.e. individual
frequencies, and stronger constraints extracted from the
observations, such as the mixed modes. We confirm the robust
determination of $\log g$ from the average seismic quantities, as our
results are within 1-$\sigma$ uncertainty of the pipeline results.

We confirmed these stars to be subgiants (having evolved off the
main sequence) and this allowed us to resolve the disagreement
between the seismic ages determined from the pipeline analyses and
the ages estimated using the lithium abundance and the empirical
Li-age relationship. Basically, the Li abundance cannot be employed
to estimate the ages of the subgiants. Similarly, existing age-rotation-activity relations can only be indicative for subgiants as these relations are calibrated mostly
for the main-sequence stars. This must be taken into account for gyrochronology studies.

We will soon obtain longer data sets from \textit{Kepler} for many more
stars and our results are a good indication of what we can achieve. We note that KIC\,10920273 and KIC\,11395018 are
at the faint end of the \textit{Kepler} asteroseismic targets;
hence, this work sets a lower limit to the quality of information we can expect from asteroseismology.

\acknowledgements \footnotesize{We thank the entire \textit{Kepler}
team, without whom these results would not be possible. Funding for
this Discovery mission is provided by NASA's Science Mission
Directorate. We also thank all funding councils and agencies that
have supported the activities of KASC Working Group 1. This article
is based on observations made with the Nordic Optical Telescope
(NOT) operated on the island of La Palma in the Spanish Observatorio
del Roque de los Muchachos. GD gratefully acknowledges financial
support from the following institutions: Advanced Study Program
(ASP) of the National Center for Atmospheric Research (NCAR), NASA
under Grant No. NNX11AE04G (together with MP and TSM), The Danish Council for
Independent Research; and thanks Y. Elsworth, S. Hekker, M. St\c{e}\'slicki, J. C. Su\'arez, M. J. Thompson, and the anonymous referee for useful comments. NCAR is partially supported by the National
Science Foundation. Funding for the Stellar Astrophysics Centre is provided by The Danish  
National Research Foundation. The research is supported in part by the  
ASTERISK project (ASTERoseismic Investigations with SONG and Kepler)  
funded by the European Research Council (Grant agreement no.: 267864). AOT acknowledges support from
Sonderforschungsbereich SFB 881 "The Milky Way System" (subproject
A5) of the German Research Foundation (DFG). IMB is supported by
the grant SFRH / BD / 41213 /2007 from FCT /MCTES, Portugal.  IMB and MJPFG were 
supported in part by grant PTDC/CTE-AST/098754/2008 from FCT-Portugal and 
FEDER. JM-\.Z acknowledges the Polish
Minstry grant number N\,N203\,405139. KB acknowledges the funding
support from the INAF Postdoctoral fellowship. This research was
carried out while OLC was a Henri Poincar\'e Fellow at the
Observatoire de la C\^ote d'Azur. The Henri Poincar\'e Fellowship is
funded by the Conseil G\'en\'eral des Alpes-Maritimes and the
Observatoire de la C\^ote d'Azur. RAG has received funding from the European Community's Seventh Framework Programme
(FP7/2007-2013) under grant agreement No. 269194 (IRSES/ASK). SGS acknowledges the support from the Funda\c{c}\~ao para a
Ci\^encia e Tecnologia (grant ref. SFRH/BPD/47611/2008) and the
European Research Council (grant ref. ERC-2009-StG-239953). TLC
acknowledges financial support from project PTDC/CTE-AST/098754/2008
funded by FCT/MCTES, Portugal. WJC acknowledges financial support
from the UK Science and Technology Facilities Council. We
acknowledge the KITP staff at UCSB for their warm hospitality during
the research program ``Asteroseismology in the Space Age".
 This research was supported in part by the National Science Foundation under Grant No. PHY05-51164.}

\bibliographystyle{aa}
\bibliography{myrefs}

\clearpage
\begin{landscape}
\begin{table}[h]
\begin{center}
\caption{Input physics used in the evolution codes} \label{input}
\vspace{12pt} {\footnotesize
\begin{tabular}{@{}lllllll}
\hline \hline \\
Team & Diffusion & Convection  & Overshoot & EOS & Opacities  & Nuclear reaction \\
 & \& settling & treatment & (core) &  & (high/low temperature) & rates \\
\hline
AMP$^{\rm a}$ (ASTEC)$^{\rm b}$& He$^{\rm c}$ & MLT$^{\rm d}$ &no&OPAL2005$^{\rm e}$&OPAL$^{\rm f}$/\citet{Alexander1994}&B \& P (1992)$^{\rm g}$\\
ASTEC1& none & MLT &no&OPAL2005&OPAL/\citet{Ferguson05}&NACRE$^{\rm h}$\\
ASTEC2& He \& heavy elements & MLT&no&OPAL2005&OPAL/\citet{Ferguson05}&NACRE\\
CESAM$^{\rm i}$ & none & CGM$^{\rm j}$ &yes&OPAL2005&OPAL/\citet{Alexander1994}&NACRE\\
Geneva$^{\rm k}$ & He \& heavy elements$^{\rm l}$& MLT &yes&OPAL2005&OPAL/\citet{Alexander1994}&NACRE\\
\hline
\end{tabular}}
\end{center}
{ \footnotesize a) \citet{metcalfe09}, b) Aarhus Stellar Evolution
Code \citep{cd08aastec}, c) as described by \citet{michaud93},
 d) Mixing length
theory \citep{vitense58}, e) \citet{rogers02},
 f) \citet{iglesias96},  g) \citet{Bahcall1992}, h)\citet{angulo99},
i) \citet{CESAM1997}, j) Canuto-Goldman-Mazzitelli model for
turbulent convection \citep{CGM1996}, k)
\citet{eggenberger2008}
l) \citet{proffitt1991}\\}
\end{table}

\begin{table}[h]
\begin{center}
\caption{Parameter space searched by each team}\label{grids} \vspace{12pt} \centering {\footnotesize
\begin{tabular}{@{}llllll}
\hline \hline \\
Team &$\rm{M/M}_{\odot}$ & $Z/X$&$Y$ & $\alpha$ &$\alpha_{\rm ov}$ \\
\hline
AMP (ASTEC) & 0.75--1.75&0.0026--0.079&0.22--0.32&$\alpha_{\rm MLT}$=1.0--3.0&N/A\\
ASTEC1&1.00--1.60 &0.01--0.07&0.24--0.32&$\alpha_{\rm MLT}$=1.8&N/A\\
ASTEC2 &1.2--1.4&0.025--0.046&0.26--0.30&$\alpha_{\rm MLT}$=1.78--1.84&N/A\\
CESAM &N/A*&0.026--0.042&0.24--0.28 &$\alpha_{\rm CGM}$ = 0.52--0.68 & 0.0--0.2\\
Geneva &1.00--1.50&0.016--0.040&0.25--0.30&$\alpha_{\rm MLT}$=1.8&0.1\\
\hline
\end{tabular}}
\end{center}
\footnotesize{ * For each given set of parameters ($Z/X, Y,
\alpha, \alpha_{\rm ov}$), the method proposed by
\citet{Deheuvels2011} results in a precise estimate of the mass by
using the observed large frequency separation
 and the frequency of the mixed modes.}
\end{table}
%\clearpage
%\end{landscape}

%\clearpage
%\begin{landscape}
\begin{table}
\begin{center}
\caption{\normalsize Fitted parameters for KIC\,10920273.} \label{scully} \vspace{12pt} 
{\scriptsize
\begin{tabular}{lcccccccccrcrr} \hline \hline \\
Model& $M/M_{\odot}$ & $(Z/X)_{\rm i}$ & $Y_{\rm i}$ & $\alpha$ &
$t$(Gyr) & $L/L_{\odot}$ & $R/R_{\odot}$ & $T_{\rm
eff}$(K) & log\,\emph{g} & [Fe/H] &$X_{\rm c}$& $\chi^{2}_{\rm seis}$&$\chi^{2}_{\rm atm}$  \\
\hline
SA1 (AMP)&1.00&0.0154&0.296&2.04&6.74& 3.31&1.779&5844&3.937& $-$0.203&0.0&6.33&2.03 \\
SA2 (AMP)& 1.03 &0.0143& 0.267& 1.98& 7.28& 3.24&1.797& 5781 &3.942&  $-$0.235&0.0& 6.35&2.10\\
SA3 (AMP)$^{\rm a}$&0.96&0.0152&0.311&1.94&6.96&3.14&1.754&5805&3.932&$-$0.208&0.0&3.09&1.90\\
SA4 (AMP)$^{\rm b}$&1.00&0.0151&0.285&2.02&6.86&3.22&1.780&5799&3.937&$-$0.210&0.0&3.60&1.85\\
SB1 (ASTEC1)&1.02&0.0200&0.290&1.80&7.59&2.97&1.788&5674&3.944& $-$0.088&0.0&11.44& 1.71\\
SB2 (ASTEC1)$^{\rm b}$&1.06&0.0300&0.300&1.80&7.45&2.84&1.813&5572&3.948&0.088&0.0&23.60&4.21\\
SC (ASTEC2)&1.10&0.0214&0.285&1.81&5.76&3.42&1.824&5819&3.960& $-$0.090&0.0&33.90&0.79\\
SD (CESAM)$^{\rm b}$&0.97&0.0152&0.300&0.56$^{\rm c}$&7.80&3.10&1.760&5779&3.930&$-$0.207&0.0&4.00&1.90\\
SE (Geneva)&1.15&0.0220&0.275&1.80&5.60&3.77&1.861&5900&3.959&$-$0.040&0.0&45.78&1.40 \\
\hline
  weighted mean       & 1.00 &0.0161&0.294&&7.12&3.18& 1.776&5787&3.937&$-$0.188\\
  standard deviation    & (0.04)&(0.0030)&(0.014)&&(0.47)&(0.13)& (0.021)&(55)&(0.007)&(0.064)\\
\hline
Creevey et al. &1.25$\pm$0.13&&&&4.5$\pm$1.8&3.6$\pm$1.2&1.90$\pm$0.05&&3.97$\pm$0.03&\\
(2012)&&&&&5.0$\pm$1.9$^{\rm d}$&&&&3.94$\pm$0.03$^{\rm e}$&\\
\hline
\end{tabular}}
\end{center}
\footnotesize{  $^{\rm a}$ Maximal-list frequencies are used as input.\\
                         $^{\rm b}$ Two frequencies are excluded from the maximal-list frequencies (see text).\\
                         $^{\rm c}$ Canuto-Goldman-Mazzitelli (CGM) model for turbulent convection \citep{CGM1996} is used in this model.\\
                         $^{\rm d}$ This is the value when the average small frequency separation is also used as a seismic constraint.\\
		       $^{\rm e}$ Asteroseismic $\log g$ obtained from scaling relations (Table 7 of \citealt{Creevey2012})}
\end{table}

\begin{table}
\begin{center}
\caption{\normalsize Fitted parameters for KIC\,11395018.}
\label{boogie} \vspace{12pt} \centering {\scriptsize
\begin{tabular}{lcccccccccrcrr} \hline \hline \\
Model& $M/M_{\odot}$ & $(Z/X)_{\rm i}$ & $Y_{\rm i}$ & $\alpha$ &
$t$(Gyr) & $L/L_{\odot}$ & $R/R_{\odot}$ & $T_{\rm
eff}$(K) & log\,\emph{g} & [Fe/H] & $X_{\rm c}$&$\chi^{2}_{\rm seis}$ & $\chi^{2}_{\rm atm}$ \\
\hline

BA1 (AMP)&1.23&0.034&0.301&1.94&4.46&4.40&2.158&5697&3.860&0.144&0.0&8.07&0.49\\
BA2 (AMP) &1.32 &0.028& 0.241& 1.94& 4.84 & 4.53 &2.210 & 5671&3.869&0.049 &0.0&8.51&0.69\\
BA3 (AMP)$^{\rm a}$&1.26&0.034&0.294&2.02&4.28&4.64&2.175&5749&3.863&0.140&0.0&7.36&0.55\\
BB1 (ASTEC1)&1.235&0.033&0.282&1.80&5.05&4.04&2.156&5573&3.861&0.114&0.0&17.03&1.02 \\
BB2 (ASTEC1)$^{\rm a}$&1.22&0.040&0.310&1.80&4.65&4.11&2.154&5605&3.858&0.213&0.0&17.08&1.02\\
BC1 (ASTEC2)&1.25&0.034&0.297&1.81&4.29&4.42&2.265&5615&3.820&0.116&0.0&47.34& 0.90\\
BC2 (ASTEC2)$^{\rm b}$&1.32&0.043&0.270&1.81&4.89&4.06&2.211&5515&3.870&0.243&0.0&17.65& 2.01\\
BD (CESAM)&1.29&0.030&0.260&0.64$^{\rm c}$&4.50&4.89&2.190&5806&3.866&0.026&0.0&4.04&1.19\\
BE (Geneva)&1.32&0.033&0.275&1.80&4.30&5.18&2.207&5867&3.871&0.120&0.0&115.69&1.37 \\
\hline
    weighted mean      & 1.27&0.033&0.276&&4.57&4.54 &2.184&5706&3.863&0.103\\
    standard dev.      & (0.04)&(0.004)&(0.022)&&(0.23)&(0.30) &(0.024)&(92)&(0.008)&(0.070)\\
\hline
Creevey et al. &1.37$\pm$0.11&&&&3.9$\pm$1.4&4.2$\pm$1.1&2.23$\pm$0.04&&3.88$\pm$0.02&\\
(2012)&&&&&4.5$\pm$0.5$^{\rm d}$&&&&3.86$\pm$0.03$^{\rm e}$&\\
\hline
\end{tabular}}
\end{center}
\footnotesize{  $^{\rm a}$ Maximal-list frequencies are used as input.\\
                         $^{\rm b}$ No diffusion is included in this model unlike other ASTEC2 models.\\
                         $^{\rm c}$ Canuto-Goldman-Mazzitelli (CGM) model for turbulent convection \citep{CGM1996} is used in this model.\\
                         $^{\rm d}$ This is the value when the average small frequency separation is also used as a seismic constraint.\\
		       $^{\rm e}$ Asteroseismic $\log g$ obtained from scaling relations (Table 7 of \citealt{Creevey2012})}
\end{table}
\clearpage
\end{landscape}

\end{document}